\documentclass[twocolumn]{aastex63}
\usepackage{multirow}
\usepackage{amssymb}
\usepackage{amsmath}

\shorttitle{HD~45364 Dynamics \& Habitability}
\shortauthors{Li et al.}

\begin{document}

\title{New Dynamical State and Habitability of the HD~45364 Planetary System}

\author[0000-0002-4860-7667]{Zhexing Li}
\affiliation{Department of Earth and Planetary Sciences, University of California, Riverside, CA 92521, USA}
\email{zli245@ucr.edu}

\author[0000-0002-7084-0529]{Stephen R. Kane}
\affiliation{Department of Earth and Planetary Sciences, University of California, Riverside, CA 92521, USA}

\author[0000-0002-4297-5506]{Paul A.\ Dalba}
\altaffiliation{Heising-Simons 51 Pegasi b Postdoctoral Fellow}
\affiliation{Department of Astronomy and Astrophysics, University of California, Santa Cruz, CA 95064, USA}

\author[0000-0001-8638-0320]{Andrew W. Howard}
\affiliation{Cahill Center for Astronomy $\&$ Astrophysics, California Institute of Technology, Pasadena, CA 91125, USA}

\author[0000-0002-0531-1073]{Howard T. Isaacson}
\affiliation{Department of Astronomy, University of California Berkeley, Berkeley, CA 94720, USA}
\affiliation{Centre for Astrophysics, University of Southern Queensland, Toowoomba, QLD, Australia}

%%%%%%%%%%%%%%%%%%%%%%%%%%%%%%%%%%%%%%%%%%%%%%%%%%%%%%%%%%%%%%%%%%%%

\begin{abstract}

Planetary systems with multiple giant planets provide important opportunities to study planetary formation and evolution. The HD~45364 system hosts two giant planets that reside within the Habitable Zone (HZ) of their host star and was the first system discovered with a 3:2 mean motion resonance (MMR). Several competing migration theories with different predictions have previously provided explanations regarding the observed resonance through dynamical simulations that utilized limited data. Here, over ten years since the original discovery, we revisit the system with a substantially increased radial velocity (RV) sample from HARPS and HIRES that significantly extend the observational baseline. We present the revised orbital solutions for the two planets using both Keplerian and dynamical models. Our RV models suggest orbits that are more circular and separated than those previously reported. As a result, predicted strong planet-planet interactions were not detected. The system dynamics were reanalyzed, and the planet pair was found to exhibit apsidal behavior of both libration and circulation, indicating a quasi-resonance state rather than being truly in MMR. The new orbital solution and dynamical state of the system confirm migration models that predicted near circular orbits as the preferred scenario. We also study the habitability prospects of this system and found that an additional Earth-mass planet and exomoons in the HZ are possible. This work showcases the importance of continued RV observation and its impact on our knowledge of the system's dynamical history. HD~45364 continues to be an interesting target for both planetary formation and habitability studies.

\end{abstract}


%%%%%%%%%%%%%%%%%%%%%%%%%%%%%%%%%%%%%%%%%%%%%%%%%%%%%%%%%%%%%%%%%%%%

\section{Introduction}
\label{intro}

Over the past decade, there has been a dramatic increase in the number of exoplanets detected, largely attributable to the \textit{Kepler} \citep{borucki2010a} and \textit{TESS} \citep{ricker2015} missions. The huge inventory of exoplanets we see today shows a vast diversity in terms of planetary intrinsic properties \citep{konacki2003b,kipping2011,barclay2013,masuda2014,gaudi2017}, orbital characteristics \citep{jones2006,smith2018,zhang2021}, and system architectures \citep{muirhead2012a,schwamb2013,cabrera2014,gillon2017a,feng2017,bohn2020}. From the over five thousand confirmed exoplanets within over three thousand planetary systems so far, there are over eight hundred systems that host multiple planets, many of which host planets in mean motion resonance (MMR) chains. MMR can occur when planets within a system have orbital periods of near integer ratios with each other. Planets that are within an MMR chain often exert significant gravitational influence on each other and exchange angular momentum periodically \citep{raymond2008,petrovich2013,goldreich2014}. Of particular interest are giant planets exhibiting MMR orbital configurations at locations within the snow line. According to standard planetary formation theories, giant planets are believed to form at distances far away from the star where temperature is low enough for condensation of volatile compounds, such as water ice \citep{sasselov2000}. Therefore, giant planets within an MMR chain at locations within the snow line are likely to have initially formed beyond the snow line, then gradually migrated inwards through convergent migration while being embedded within the protoplanetary disk. Such interacting planets undergoing migration thus provide important constraints on planetary formation, planet migration history, system evolution and architecture \citep{correia2009,rosenthal2019}. Additionally, the presence of giant planets within the system could greatly influence the habitability of terrestrial planets \citep{georgakarakos2018,sanchez2018,kane2019e,kane2020b,bailey2022b}. Giant planets passing through the system's habitable zone (HZ) during migration provides important clues to the viability of forming additional terrestrial planets in the HZ around the host star, and if so, the dynamical stability of these low mass planets. Systems such as these with giant planets interacting with each other in MMR, particularly in the HZ of the star, offer rare opportunities to simultaneously study the intriguing dynamical history of the planets as well as habitability of the system.

The star HD~45364 was discovered to host two gas giant planets by \citet[C09 hereafter]{correia2009} using radial velocity (RV) data from the High Accuracy Radial velocity Planet Searcher (HARPS) spectrograph \citep{pepe2000}. The two planets were found to be trapped in a 3:2 MMR orbiting inside the HZ of the system, and was the first exoplanetary system found to host planets exhibiting such an orbital configuration. Orbital parameters published by C09 predicted strong planet-planet interactions and the interesting dynamics of the system prompted several studies into the possible planetary formation and migration route that could explain the observed MMR. Notably, \citet[R10 hereafter]{rein2010} proposed a formation scenario where the outer planet underwent a very rapid inward type III migration \citep{masset2003} that allowed it to quickly pass through the very stable 2:1 resonance and finally settle in a 3:2 resonance with the inner planet. This scenario was verified by R10 through both N-body and hydrodynamical simulations under the assumptions that the surface density of the protoplanetary disk was about five times higher than the minimum solar nebula \citep[MMSN;][]{hayashi1981} at 1 au and that there was no active mass accretion during the process of migration. That is, the two planets achieved their final orbital states with giant planet masses before the start of the migration. Instead of the unconventional type III migration model that R10 proposed, \citet[C13 hereafter]{correa-otto2013} suggested that a more traditional model with the combination of type I and type II migration is possible to explain the observed 3:2 MMR. In this scenario, both planets were assumed to be in their planet embryo states initially with masses of 0.1~$M_\oplus$ and started the migration under the type I regime within the disk with surface density equivalent to MMSN at 5~au. The two planets were allowed to accrete mass while migrating inward under the two-stage giant planet formation model by \citet{alibert2005}, where migration effects are included during the formation. The planet pair was able to pass through 2:1 and other MMRs due to their small masses and minimal interaction with each other. This scenario by C13 suggests that the two planets were able to reach near the 3:2 MMR at the end of the first stage of planetary formation before reaching their critical masses and the onset of runaway growth. Both planets would then be massive enough to continue migration towards the observed orbital configuration slowly under type II migration.

One of the intriguing aspects about the different proposed migration scenarios of the HD~45364 planet pair is that they predicted different orbital configuration outcomes. The outer planet type III migration to a 3:2 resonance capture with the inner planet modeled in R10 resulted in much lower-eccentricity, near circular orbits for both planets and, as a result, different libration patterns from that reported in C09. With the planet pair undergoing type I and II migration in C13, two types of simulations were conducted, K3 and K100, with different ratios (3 and 100) of the e-folding times of semi-major axis decay and eccentricity damping for both planets assumed for their type II migration stage. Interestingly, the K100 simulation reproduced a similar result to that in R10 while K3 simulation was consistent with the configuration originally derived in C09. It is worth noting that at the time of these publications, RV data were limited to an observation baseline of $\sim$4~years; perhaps too short to detect the longer term effect of the system dynamics and distinguish between the different proposed models. However, the previous works have stressed that the effort to resolve the orbital configuration and migration model degeneracies could greatly benefit from further RV observations.

In this work, we provide a clearer picture of the orbital configuration of the two planets as well as the system architecture using the latest RV dataset. This dataset, as will be described in Section~\ref{rvsol}, doubles the amount of RV data and quadruples the observation baseline compared to the previously published data. Such a data expansion allows us to revisit this fascinating system after over ten years to refine the orbital parameters for both planets and attempt to detect the predicted strong planet-planet interaction through both Keplerian and dynamical modeling of the RV data. We also provide constraints on the system's orbital inclination using only RV information by fitting the data with dynamical models at different inclinations. An RV survey completeness analysis is conducted and we explore the potential of additional undetected planets within the system. The new orbital configuration has particular implications for the dynamical state of the system, which we also study in detail. We are especially interested in the possibility of detecting additional low mass terrestrial planets in the HZ of the system. As mentioned earlier, the presence of two giant planets in the HZ as a result of inward migration may have a huge impact on the formation of other, smaller planets in the HZ. The confirmation or exclusion of possible additional low mass planets in the system could provide a clue to the habitability prospects of this system and will be useful for target selection processes related to future missions that search for potentially habitable terrestrial planets.

The paper is organized as follows: in Section~\ref{system}, we describe the HD~45364 system architecture and the extent of the HZ. We present the revised RV solutions of the system with the new RV data collected using both Keplerian and dynamical models in Section~\ref{rvsol}, including orbital inclination constraints and RV completeness calculations. Section~\ref{dynamics} discusses an analysis of the system's new dynamical state near MMR, and the planet-planet interaction within that resonance regime. Prospects for terrestrial bodies within the HZ of the system, including planets and exomoons, are investigated in Section~\ref{habitability}. We finally discuss the implications of the results from this work and provide concluding remarks in Section~\ref{conclusions}.

%%%%%%%%%%%%%%%%%%%%%%%%%%%%%%%%%%%%%%%%%%%%%%%%%%%%%%%%%%%%%%%%%%%%

\section{System Architecture and Habitable Zone}
\label{system}

\begin{figure*}[htbp!]
    \begin{center}
        \begin{tabular}{cc}
            \includegraphics[width=8.5cm]{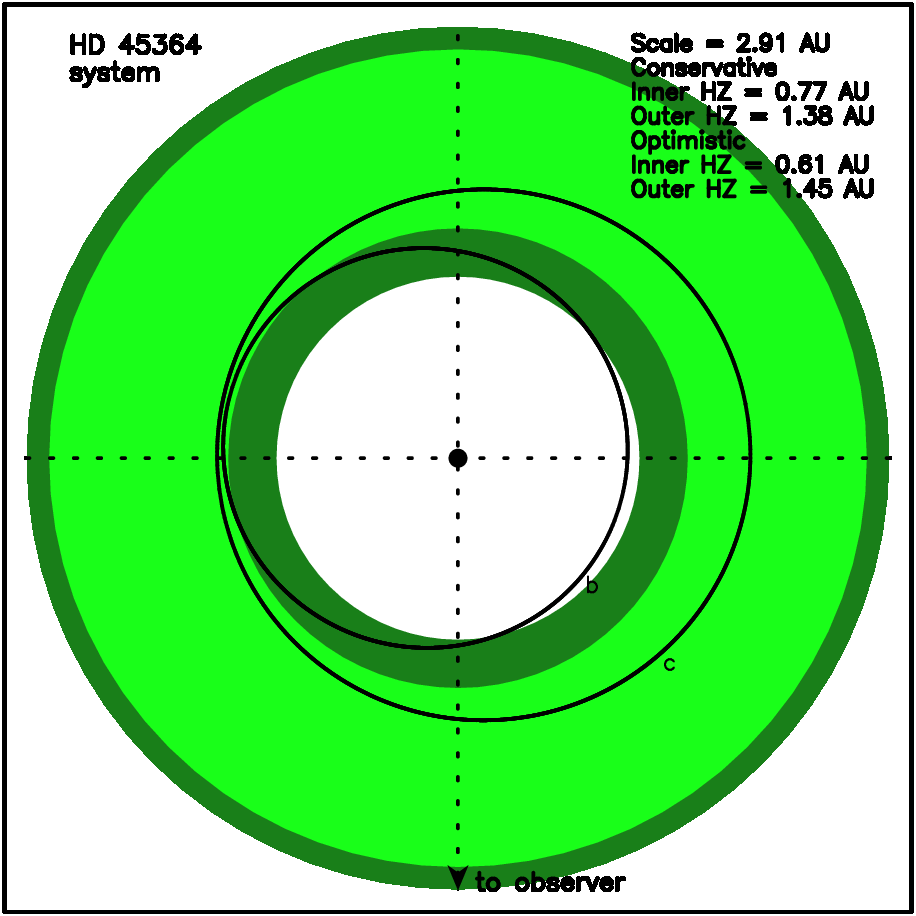} &
            \includegraphics[width=8.5cm]{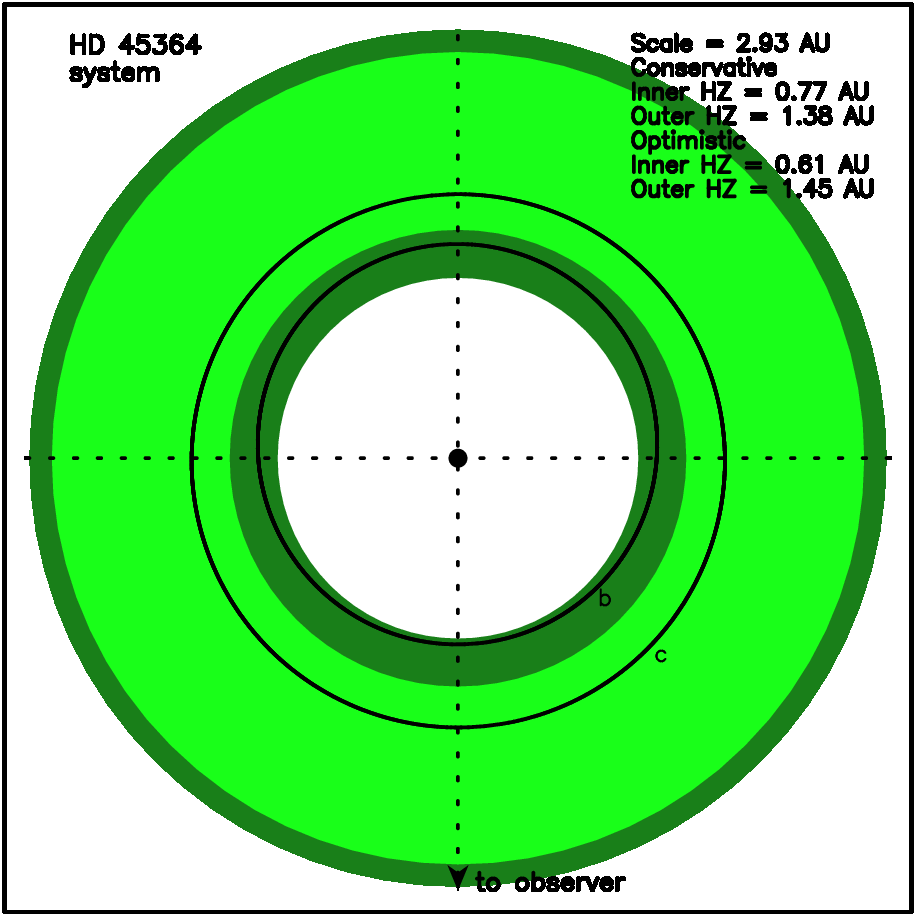}
        \end{tabular}
    \end{center}
  \caption{A top-down view of the planetary system HD~45364, showing the orbits of the b and c planets, where the left panel shows those reported by C09. The green annuli represent the extent of the HZ for the system, where the light green indicates the extent of the CHZ, and the dark green indicates the OHZ extension to the CHZ. The right panel shows the revised orbits with the new RV data presented later in this paper.}
  \label{fig:hzorbits}
\end{figure*}

HD~45364 is a nearby star with a V band magnitude of 8.08 \citep{correia2009}, and is located at a distance of 34.35~$\pm$~0.04~pc \citep{gaia2018}. The star has a spectral type K0V, an effective temperature of 5466$^{+59}_{-32}$~K, and a luminosity of 0.637 $\pm$ 0.002~$L_\odot$ \citep{gaia2018}. The mass and radius were estimated to be 0.82~$\pm$~0.05~$M_\odot$ and 0.89$^{+0.01}_{-0.02}$~$R_\odot$, respectively \citep{correia2009,gaia2018}. The system was reported by C09 to host two giant planets through the HARPS RV search with minimum masses of 0.1872~$M_{\rm J}$ and 0.6579~$M_{\rm J}$ for the b and c planets, respectively. The orbits of the b and c planets have semi-major axes of 0.6813~au and 0.8972~au, respectively, and eccentricities of 0.168~$\pm$~0.019 and 0.097~$\pm$~0.012, respectively. Error estimates unfortunately were not provided by C09 for both minimum masses and semi-major axes. These reported orbits are summarized in Table~\ref{tab:param} and depicted in a top-down view of the system, shown in the left panel of Figure~\ref{fig:hzorbits}, showing the proximity of the two planetary orbits that potentially render the system unstable. However, dynamical simulations performed by C09 found that the two planets, with orbital periods 226.93~$\pm$~0.37~days and 342.85~$\pm$~0.28~days, reside within a 3:2 MMR such that the planets never pass within 0.37~au of each other. The system configuration was thus determined to be stable over Gyr timescales, making it the first detected exoplanetary system to exhibit such a MMR and an interesting case study of planetary formation and orbital dynamical scenarios.

In addition to the MMR configuration, the locations of the two planets are intriguing to the study of planetary habitability in the habitable zone \citep[HZ;][]{kopparapu2013a,kopparapu2014,kane2016c}, and also for the study of potentially habitable exomoons \citep{heller2014c,hill2018}. We calculated the boundaries of the conservative and optimistic HZ (CHZ and OHZ, respectively) in the system following the definitions described by \citet{kopparapu2013a,kopparapu2014}. We adopted the aforementioned stellar luminosity and effective temperature values. The inner and outer boundaries for the CHZ were determined to be 0.77~au and 1.38~au, respectively. For the OHZ, the inner and outer boundaries were found to be 0.61~au and 1.45~au, respectively. The extent of these HZ boundaries are depicted in Figure~\ref{fig:hzorbits}, along with the C09 planetary orbits described above. The left panel of Figure~\ref{fig:hzorbits} demonstrates that the orbit of planet c lies completely within the bounds of the CHZ, whilst planet b spends 86\% of its orbital period within the CHZ and the remaining time either in the inner OHZ region or interior to the HZ entirely. Although giant planets are likely poor considerations for habitable conditions, additional terrestrial exoplanets within the HZ or exomoons orbiting the known giant planets could potentially have feasible environments for life \citep{heller2014c}. The right panel of Figure~\ref{fig:hzorbits} shows a depiction of the revised orbital solution for both planets from this work, and is described in detail in Section~\ref{rvsol}.

%%%%%%%%%%%%%%%%%%%%%%%%%%%%%%%%%%%%%%%%%%%%%%%%%%%%%%%%%%%%%%%%%%%%

\section{Revised Radial Velocity Solution}
\label{rvsol}

The original discovery of the two-planet system by C09 utilized 58 RVs acquired using the HARPS spectrograph \citep{pepe2000} on the ESO 3.6-m telescope at La Silla, Chile in the span of 1583 days from December 2003 to April 2008. The best-fit orbital solution at that time (see Table~\ref{tab:param} and Figure~\ref{fig:hzorbits}) places the two planets in a 3:2 MMR and slightly eccentric orbits with the potential for close encounters. Since then, HD~45364 had been continuously monitored by the HARPS team for an extended period of time until September 2017, extending the RV baseline to over 5000 days with 114 data points in total. The full HARPS dataset for this target was later re-reduced with the new SpEctrum Radial Velocity AnaLyser (SERVAL) pipeline \citep{zechmeister2018} and had several systematics corrected for including nightly zero-point variations and intra-night drifts. The newly reduced HARPS data achieved slightly higher precision and was published as part of the HARPS RV database \citep{trifonov2020}. The fiber upgrade to the HARPS instrument in May 2015 \citep{locurto2015} modified the instrument profile and created a vertical offset between the pre- and post-upgraded RV data \citep{trifonov2020}. Because of the upgrade, we treated the pre- and post-upgraded HARPS RV as two separate instruments during the fitting process. In addition to the HARPS dataset, we obtained 7 RVs taken using the HIRES spectrometer \citep{vogt1994} on Keck I at Mauna Kea, Hawaii from December 2009 to September 2021 (Table \ref{tab:HIRES}). The data reduction follows the similar procedure as in \citet{howard2010a}. In total, 121 RV data points were collected for this system with a baseline about 6500 days. The improved RV precision as well as the extended observational baseline, 5000 days longer than the original dataset, allow the opportunity to revisit and provide a more accurate orbital solution to the two planets, as suggested by the original paper.

\begin{deluxetable}{lcr}[htbp]
    \tablecaption{HIRES RV Measurements of HD~45364.
    \label{tab:HIRES}}
    \tablewidth{\columnwidth}
    \tablehead{
        \colhead{Time (BJD - 2,450,000)} &
        \colhead{RV (m s$^{-1}$)} &
        \colhead{$\sigma$ (m s$^{-1}$)}}
    \startdata
    5,167.004791 & -19.13 & 1.10 \\ 
    5,201.024304 & -6.61 & 1.04 \\ 
    5,257.889444	& 23.10 & 1.06  \\
    9,118.139321	& 28.48 & 1.03 \\ 
    9,208.913846	& -11.01 & 1.03 \\ 
    9,238.836781 & -26.80 & 1.18 \\ 
    9,483.121358 & 8.88 & 1.00 \\ \hline
    \enddata
\end{deluxetable}
%%%%%%%%%%%%%%%%%%%%%%%%%%%%%%%%%%%%%%%%%%%%%%%%%%%%%%%%%%%%%%%%%%%%

\subsection{Keplerian Model}
\label{kepsol}

Given the extended baseline and additional observations obtained since the original publication, we conducted a fourier analysis for the entire dataset to check for possible additional periodicities or linear trends that may be indicative of additional companions in the system. We used \texttt{RVSearch} \citep{rosenthal2021}, an iterative planet searching tool that uses the change in the Bayesian Information Criterion ($\Delta$BIC) between models as a measure of the goodness of the fit. We set a period grid from 2 to 10,000 days and allowed linear trends and curvatures to be incorporated in the period search. Signals were considered significant if they peak above the 0.1\% false-alarm-probability (FAP) level. Two signals with periods similar to those of the originally reported planets were present and were picked up by the periodogram as expected, but with no long term linear or quadratic trend detected.

\begin{figure}[htbp!]
  \includegraphics[trim=20 20 20 20,clip,width=\columnwidth]{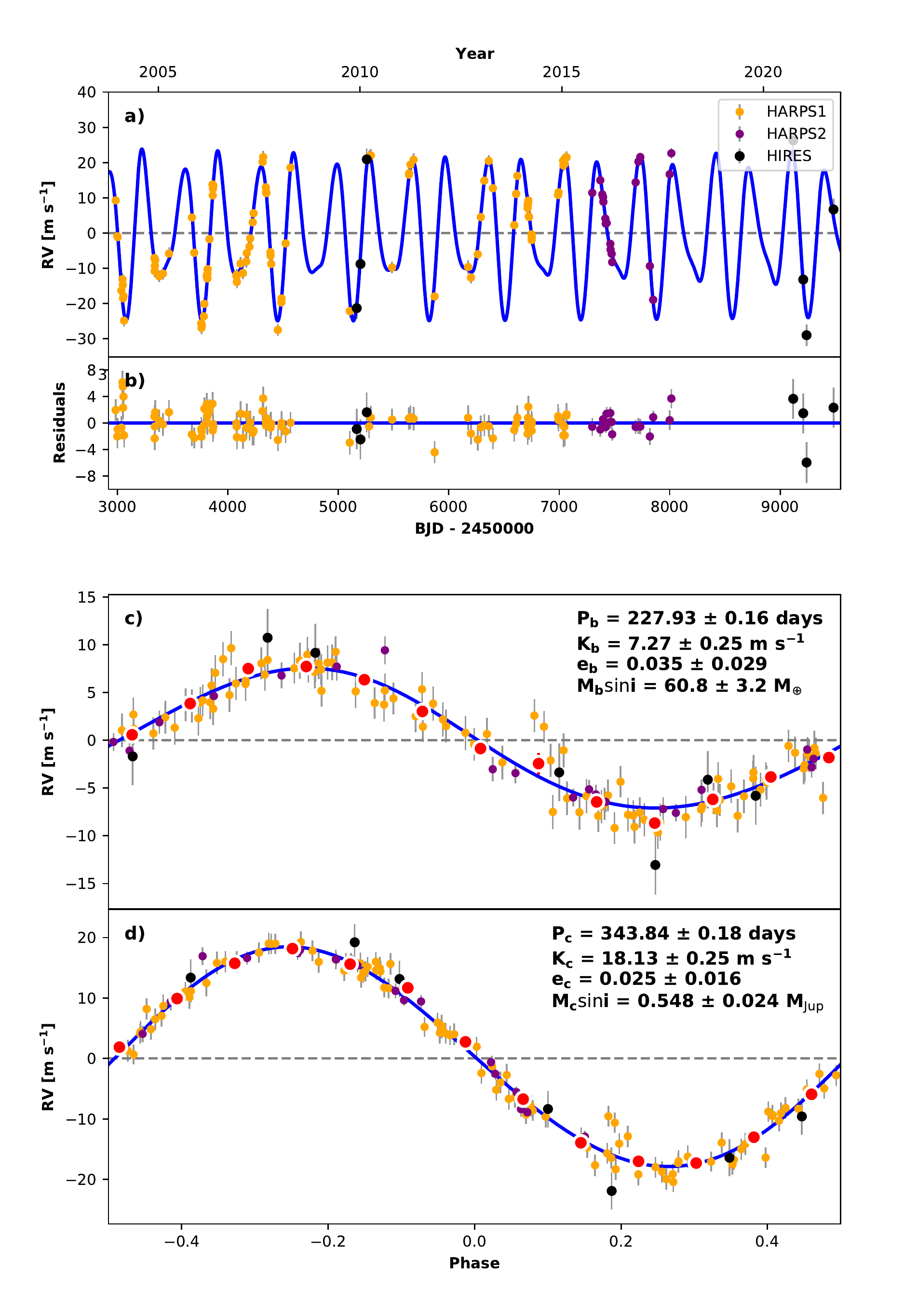}
  \caption{Best-fit Keplerian model to the RV data from pre- and post-upgraded HARPS (HARPS1 and HARPS2, respectively), and HIRES spectrographs. Best-fit model is in blue with data color coded according to instruments. Upper panels shows the total RV model and the residual after the subtraction of the best-fit model from the data. Lower panels show the phase-folded fit for each of the two planets.}
  \label{fig:radvel}
\end{figure}

We then sampled model posteriors using the RV modeling toolkit \texttt{RadVel} \citep{fulton2018a}. We used five parameters: orbital period $P$, time of inferior conjunction $T_{c}$, $\sqrt{e}$cos$\omega$, $\sqrt{e}$sin$\omega$, where $e$ and $\omega$ are orbital eccentricity and argument of periastron, respectively, and RV semi-amplitude $K$\ as the fitting basis to better sample near circular orbits and to minimize the bias towards higher orbital eccentricities during the fitting process \citep{lucy1971}. A Markov Chain Monte Carlo (MCMC) search of the parameters' posterior space was carried out using parameters returned from \texttt{RVSearch} as the initial guess. All parameters including instrumental offset and jitter terms were allowed to explore freely with no additional constraint other than forcing positive RV semi-amplitude ($K$) values, eccentricities ($e$) to be between 0 and 1, and jitter terms between 0 and 10~m~s$^{-1}$. The chain converged relatively quickly, and Figure~\ref{fig:radvel} shows the best-fit model along with the data used. The model returned approximately similar result as reported by the original paper, except in this case, orbital eccentricities for both planets were favored to be near circular, as supported by model comparison, in contrary to the mild eccentricities of $e$~$\sim$~0.17 and $e$~$\sim$~0.1 for planet b and c, respectively, that were originally reported in C09. In addition, the derived mass of the outer planet is slightly smaller than the previously reported value. As a consequence of more circular orbits, the fitted argument of periastron ($\omega$) and time of periastron ($T_{p}$) values were not as well constrained as those reported in C09. The fit produces a log-likelihood ($\ln$~$\mathcal{L}$)~=~-240.67, rms~=~1.76~m~s$^{-1}$, and $\chi_{\nu}^{2}$~=~1.15. We present the Keplerian model fit results including solution with 16\%, 50\%, and 84\% quantiles as well as that for the maximum likelihood in Table~\ref{tab:param}. We also show results from C09 in the same table for comparison.

%%%%%%%%%%%%%%%%%%%%%%%%%%%%%%%%%%%%%%%%%%%%%%%%%%%%%%%%%%%%%%%%%%%%

\subsection{Dynamical Model}
\label{dynasol}

Pure Keplerian model assumes planets orbiting around their host stars are unperturbed by other external forces other than that from the central star. However, there were indications from previous works that, due to the close proximity of the two planets with high minimum masses in MMR, strong planet-planet interactions are expected \citep{correia2009}. RVs of planets undergoing strong interactions may gradually deviate from the true Keplerian model and such deviations can be detected in the RV dataset if the observational baseline is long enough. Parameters derived using a Keplerian model would thus become an inaccurate representation of the system of interest. To account for the potential presence of planet-planet interaction between the b and c planet, a full dynamical model is needed to integrate the planets along their orbits and obtain the induced motion on the host star at each step.

We made use of the code \texttt{RVStab} \citep{rosenthal2019} as the backbone of our analysis to dynamically fit the RV data by integrating the entire system with the inclusion of planet-planet interaction using an N-body integrator \texttt{REBOUND} \citep{rein2012a}, and to carry out posterior parameter search with MCMC exploration. We used IAS15 \citep{rein2015a}, a non-symplectic integrator with adaptive time stepping, and chose the coordinate system to be with reference to the central star. We assumed coplanarity for the system and tested the case where the system inclination is edge-on ($i$~=~90$^\circ$) first. The dynamical model for the system was set up using a set of five parameters of orbital period $P$, planetary mass $M_{p}$, planet's mean longitude $\lambda$ (referenced to the first data point in our dataset), Lagrange orbital elements $e$sin$\varpi$, and $e$cos$\varpi$ for each planetary body with the addition of three jitter terms for the three instruments. $\varpi$ is longitude of periastron which is defined as $\varpi$ = $\omega$ + $\Omega$ with $\Omega$ being longitude of ascending node, which is assumed to be zero in our model. We used the results from the Keplerian fit (Table~\ref{tab:param}) and introduced a small Gaussian deviation from the parameters as an initial guess to initialize the MCMC walkers. At each step, the set of orbital parameters were fitted to the RV data and a $\ln$~$\mathcal{L}$ value was calculated for each instrument using Equation \ref{eqn:likelihood} below and were summed for all instruments as a measure of the goodness of the fit with the parameters at the current step:

\begin{multline}
    \label{eqn:likelihood}
    \ln \mathcal{L} = -\sum_{i}^{N}\frac{(\upsilon_{o,i} - \upsilon_{\mathrm{off}} - \upsilon_{m,i})^{2}}{2(\sigma_{i}^{2} + \sigma_{\mathrm{jit}}^{2})} \\- \sum_{i}^{N}\ln(\sqrt{2\pi(\sigma_{i}^{2}+\sigma_{\mathrm{jit}}^{2})}) + \ln(\sqrt{2\pi\sigma_{z}^{2}})
\end{multline}

where $\upsilon_{o,i}$, $\upsilon_{\mathrm{off}}$, and $\upsilon_{m,i}$ are $i$th observed velocity, instrumental velocity offset, and model velocity, respectively. $\sigma_{i}$ is $i$th measurement error associated with $\upsilon_{o,i}$; $\sigma_{\mathrm{jit}}$ is estimated jitter for the instrument; and $\sigma_{z}^{2}$ is defined as:

\begin{equation}
    \label{eqn:sigmaz}
    \sigma_{z}^{2} = (\sum_{i}^{N}\frac{1}{\sigma_{i}^{2}+\sigma_{\mathrm{jit}}^{2}})^{-1}
\end{equation}

\begin{figure}[htbp!]
  \includegraphics[trim=60 0 60 30,clip,width=\columnwidth]{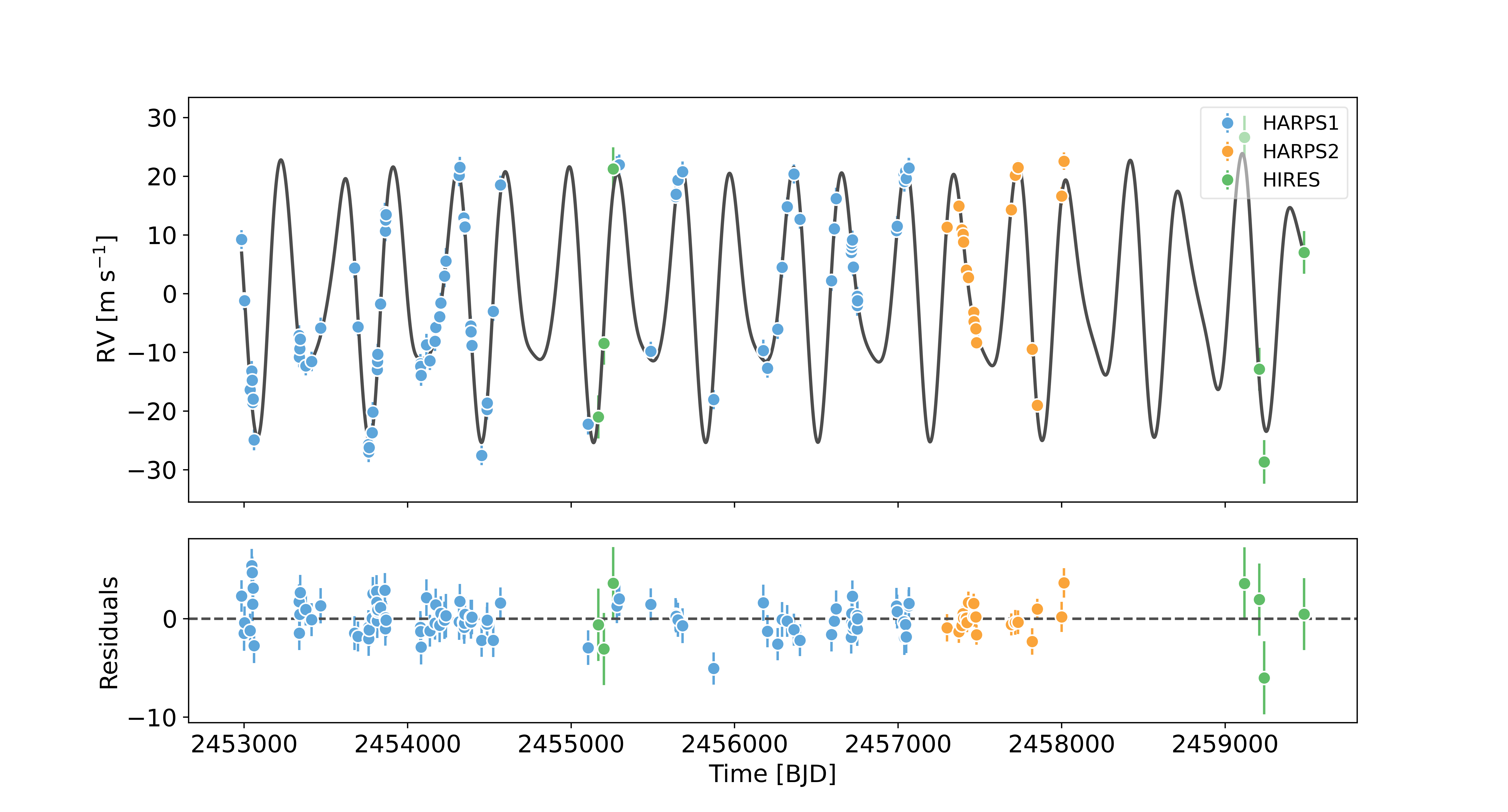}
  \caption{Best-fit dynamical model to the same RV data shown in Figure~\ref{fig:radvel}. Best-fit model is in black.}
  \label{fig:dynarv}
\end{figure}

The walkers were able to explore freely under uniform priors between the specified lower and upper bounds of fitted parameters: between 100 and 500 days for $P$; 0.0157 and 5 Jupiter masses for $M_{p}$; 0 and 360 degrees for $\lambda$; -1 and 1 for $e$sin$\varpi$ and $e$cos$\varpi$; and between 0 and 10 for the jitters. The chains were considered converged once the length of the chain was at least 50 times the autocorrelation time and the change in autocorrelation time between steps was smaller than 1\% for five consecutive steps. The chain was able to converge successfully and the fit is shown in Figure~\ref{fig:dynarv}. The orbital solution with 16\%, 50\%, and 84\% quantiles as well as maximum likelihood values are reported in Table~\ref{tab:param}. The orbital parameters from the dynamical fit are largely consistent with the one from the Keplerian fit, where we see much more circular orbits for both the b and c planet compared to the solution from C09. The mass of the outer planet appears to be smaller as well. However, we do notice the difference between the fitted orbital periods of the two planets from the dynamical model and the Keplerian model, possibly indicating the presence of planet-planet interaction. The difference in orbital periods derived from the Keplerian and dynamical models are 3-sigma significant for both planets if quoting period uncertainty values from the dynamical model. If using uncertainties from the Keplerian model, the period difference amounts to 13 and 9 sigmas for planet b and c, respectively. The dynamical model with edge-on inclination yielded $\ln$~$\mathcal{L}$~=~-232.49, rms~=~1.75~m~s$^{-1}$, and $\chi_{\nu}^{2}$~=~1.12. The dynamical model appears to be a slightly better fit to the RV data than the Keplerian model from Section \ref{kepsol} based on $\ln$~$\mathcal{L}$.

\begin{figure}[htbp!]
  \includegraphics[trim=40 0 60 30,clip,width=\columnwidth]{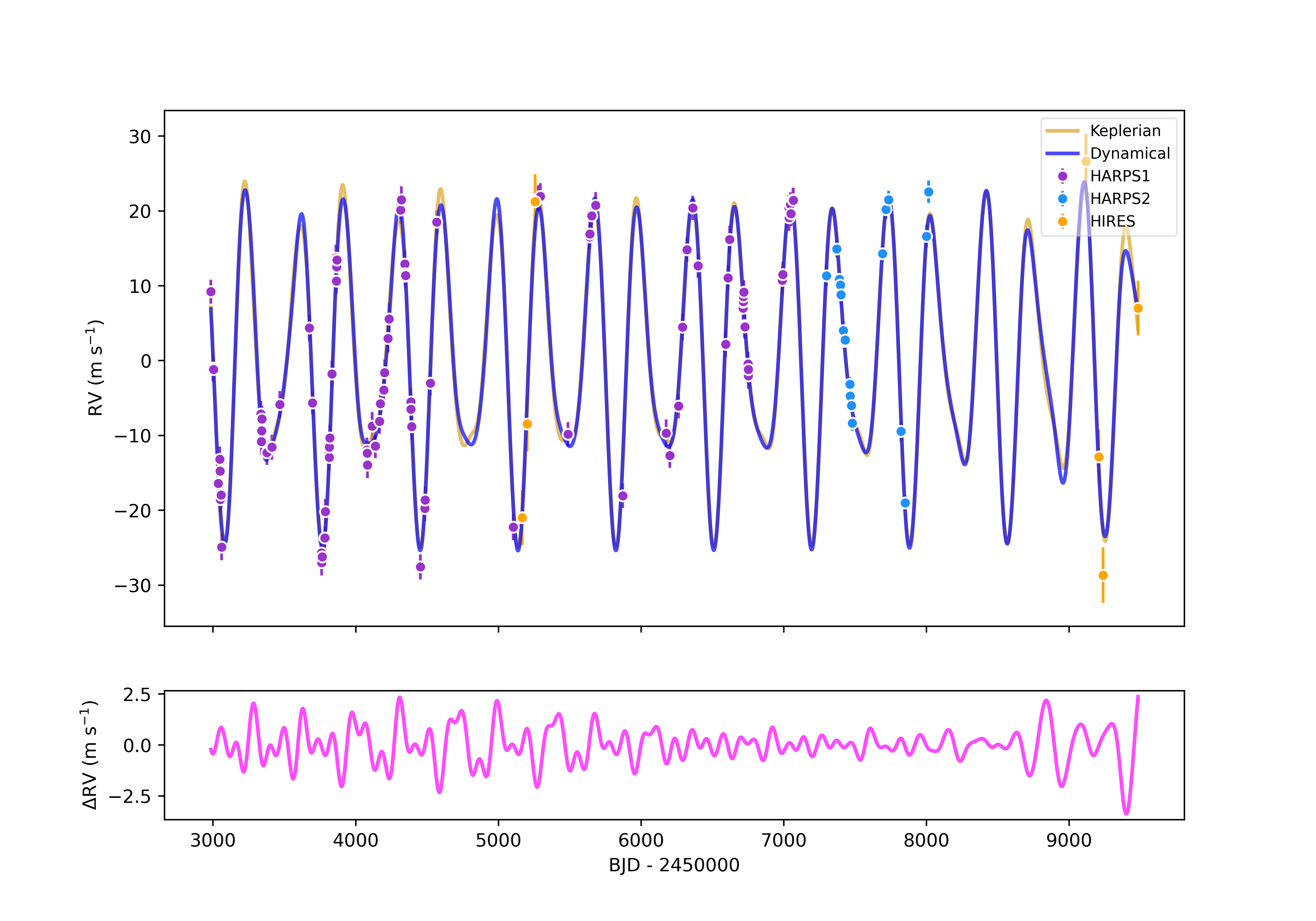}
  \caption{RV difference between the best-fit Keplerian and dynamical models in this work with the same data from Figure~\ref{fig:radvel} and \ref{fig:dynarv}. In the upper panel, Keplerian model is in gold and dynamical model is in blue. $\Delta$RV between the two models is shown in purple in the bottom panel.}
  \label{fig:deltarv}
\end{figure}

In order to detect the presence of strong planet-planet interaction claimed by previous works, we computed time series RV difference ($\Delta$RV) between the two models and the result is shown is Figure \ref{fig:deltarv}. It was predicted based on the limited data at the time of discovery in Figure~2 of C09, that the interaction between the two planets would start to become detectable in the RV data around 2015, and the $\Delta$RV between Keplerian and dynamical models would then gradually increase to an amplitude of around 15~m~s$^{-1}$. However, as can be seen in the bottom panel of Figure~\ref{fig:deltarv}, the $\Delta$RV between the two models is fluctuating on a similar level over time with low variations from sub-1 to 2.5~m~s$^{-1}$. This value is consistent with the estimated stellar jitter of $\sim$~2.34 m~s$^{-1}$ that we calculated for the host star HD~45364 following the methodology in \citet{isaacson2010}. Even at the time of our last data point, which is taken by HIRES near the end of September, 2021, $\Delta$RV does not become any more significant. Therefore, we conclude that, contrary to the previous prediction, interaction between the two planets is not as strong as expected and is in fact negligible within the sensitivity of current RV dataset. The non-detection of planet-planet interaction is possibly due to the more circular orbits and a lower mass for the outer planet derived from the new RV data in this work.

\begin{deluxetable*}{ccccccccc}[htbp!]
    \tablecaption{Orbital parameters of the two planets in HD~45364 from C09 and this work.
    \label{tab:param}}
    \tablehead{
        \colhead{Source} & 
        \colhead{Planet} &
        \colhead{$P$ (days)} &
        \colhead{$e$} &
        \colhead{$\omega$ (deg)} &
        \colhead{$T_{p}$ (BJD)} &
        \colhead{$K$ (m~s$^{-1}$)} &
        \colhead{$a$ (au)} &
        \colhead{$M_{p}$ ($M_{\rm J}$)}
    }
    \startdata
    \multirow{2}{*}{C09} & b & $226.93 \pm 0.37$ & $0.168 \pm 0.019$ & $162.6 \pm 6.3$ & $2,453,464 \pm 4$ & $7.22 \pm 0.14$ & 0.6813 & 0.187 \\
     & c & $342.85 \pm 0.28$ & $0.097 \pm 0.012$ & $7.4 \pm 4.3$ & $2,453,407 \pm 4$ & $21.92 \pm 0.43$ & 0.8972 & 0.658 \\
    \hline
    \multirow{2}{*}{\shortstack{This Work (K) \\ Quantiles}} & b & $227.93^{+0.16}_{-0.17}$ & $0.035^{+0.033}_{-0.024}$ & $232^{+63}_{-80}$ & $2,453,471^{+53}_{-35}$ & $7.27^{+0.25}_{-0.24}$ & $0.684 \pm 0.014$ & $0.191 \pm 0.010$\\
     & c & $343.84 \pm 0.18$ & $0.025^{+0.016}_{-0.015}$ & $123^{+36}_{-31}$ & $2,453,200^{+180}_{-25}$ & $18.13 \pm 0.25$ & $0.899^{+0.018}_{-0.019}$ & $0.548^{+0.023}_{-0.024}$\\
    \hline
    \multirow{2}{*}{\shortstack{This Work (K) \\ Max Likelihood}} & b & 227.95 & 0.051 & 232 & 2,453,462 & 7.32 & 0.662 & 0.180 \\
     & c & 343.93 & 0.033 & 120 & 2,453,186 & 18.17 & 0.871 & 0.513\\
     \hline
     \multirow{2}{*}{\shortstack{This Work (D) \\ Quantiles}} & b & $225.79^{+0.81}_{-0.76}$ & $0.067 \pm 0.016$ &
     $92^{+22}_{-25}$ & $2,453,374^{+15}_{-16}$ & $7.23 \pm 0.24$ & $0.6793^{+0.0016}_{-0.0015}$ & $0.1893^{+0.0062}_{-0.0063}$ \\
     & c & $345.43^{+0.54}_{-0.57}$ & $0.019^{+0.011}_{-0.010}$ & $244^{+54}_{-68}$ & $2,453,306^{+52}_{-65}$ & $18.15^{+0.25}_{-0.24}$ & $0.9020 \pm 0.0010$ & $0.5490^{+0.0075}_{-0.0074}$ \\
     \hline
     \multirow{2}{*}{\shortstack{This Work (D)* \\ Max Likelihood}} & b &
     225.34 & 0.070 & 92 & 2,453,375 & 7.26 & 0.6784 & 0.1897 \\
     & c & 345.76 & 0.010 & 276 & 2,453,336 & 18.17 & 0.9026 & 0.5497
    \enddata
    \tablecomments{Letter K and D in the Source column denote Keplerian and dynamical model from this work, respectively. $\omega$ values reported here are those of the planets, not that of the star as usually reported in RV discoveries. $T_{p}$ values for C09 were derived using $\lambda$ values reported in Table~2 of C09. For the parameters from the dynamical model in this work, $e$ and $\omega$ were derived from fitted $e$sin$\varpi$ and $e$cos$\varpi$ ($\omega$ was used instead of $\varpi$ for the Keplerian model since RV data contain no information regarding $\Omega$. $\Omega$ was assumed to be 0 for the dynamical fit.); $T_{p}$, $K$, and $a$ were derived from $\lambda$, $M_{p}$, and $P$, respectively. All models in this table assume edge-on orbital inclinations. The asterisk denotes the best-fit model we employ for the rest of this work.}
\end{deluxetable*}

%%%%%%%%%%%%%%%%%%%%%%%%%%%%%%%%%%%%%%%%%%%%%%%%%%%%%%%%%%%%%%%%%%%%

\subsection{Inclination Constraint}
\label{incli}

Typical RVs are insensitive to orbital inclination information so that systems observed with RV data have only minimum masses $M_{p}$sin$i$, instead of true masses of planets reported. However, when planets are massive enough and are orbiting close to each other, the potential dynamical interaction between the planets could provide a rare opportunity for system inclination to be constrained. This can be done by introducing orbital inclination as an additional variable and run the dynamical fit under different inclination assumptions. Since varying the assumed inclination changes the true planetary mass values, planets would start to strongly interact with each other at some inclination angles. The interaction at those angles would become so much, such that the system would be in unstable states or the resulting RV contribution from the planets would deviate from the observed RV data points, allowing us to rule out inclination cases when these situations happen. In some situations where planets are interacting strongly, orbital inclination of the system could be identified with high certainty using RV data alone. In others where no planet-planet interaction can be detected, inclination can still be constrained to within a certain range.

In this case, we first varied the system's inclination, and the masses of two planets accordingly, from edge-on to face-on to test the range of inclinations where the system would become unstable. We assumed coplanarity for the planets and used orbital solutions that assumed edge-on case derived from the dynamical work as presented in Section~\ref{dynasol} and Table~\ref{tab:param}. Once again, we employed \texttt{REBOUND} to carry out the dynamical integration using a symplectic integrator WHFast \citep{rein2015c} with an integration time step of $\sim$5.5 days, 1/40th of the b planet's orbital period and half of the recommended step size from \citet{duncan1998} to ensure proper orbital sampling. The system was integrated for $10^7$~years for each inclination angle and system stability breaks down at angles smaller than 9$^{\circ}$. 

\begin{figure}[htbp!]
  \includegraphics[trim=0 0 40 30,clip,width=\columnwidth]{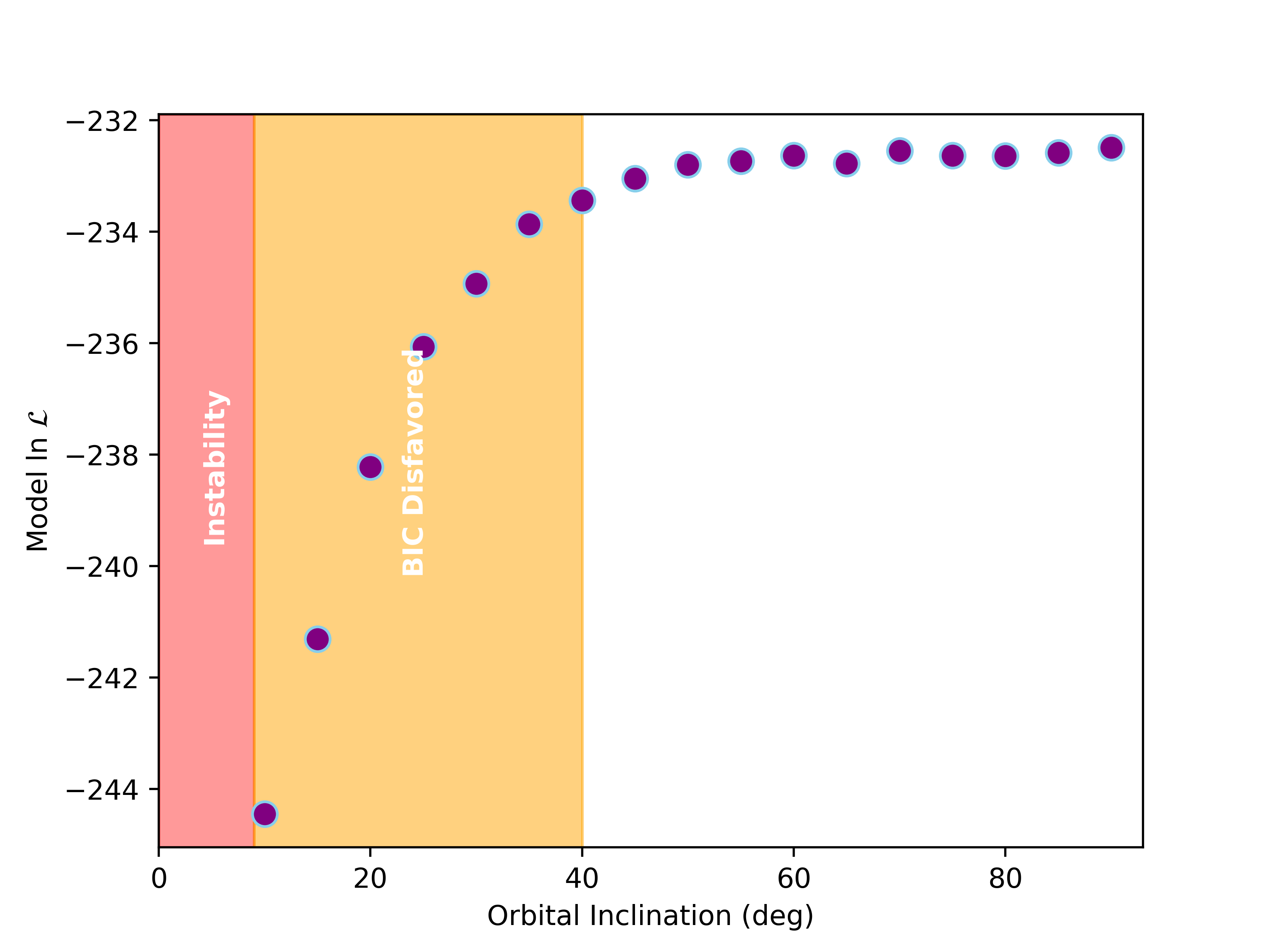}
  \caption{$\ln$~$\mathcal{L}$ for models with different inclination cases, where 0$^{\circ}$ and 90$^{\circ}$ represent face-on and edge-on orbits, respectively. Inclination step size is 5 degrees. The system becomes unstable for inclination angles smaller than 9$^{\circ}$ and angles up to 40$^{\circ}$ are disfavored by BIC.}
  \label{fig:inclikelihood}
\end{figure}

Next, we carried out similar dynamical fits to the RV data as presented earlier, except this time, dynamical models at different inclination values up to the instability angle are tested against the RV data, rather than just the edge-on case. Inclination angles were varied every 5$^{\circ}$ and the $\ln$~$\mathcal{L}$ value was recorded for each model. In the end, BIC values were calculated for all models as a comparison of goodness-of-fit. The result is shown in Figure~\ref{fig:inclikelihood}. Inclination angles leading to system instability are shaded in red, and angles with $\Delta$BIC~$>$~2 compared to the best-fit model are shaded in orange and are disfavored. Although there is no distinct peak that would allow us to pinpoint the actual inclination value, the inclination of the system could still be constrained to $\geq$~40$^{\circ}$ given models with inclinations between 40$^{\circ}$ and 90$^{\circ}$ are indistinguishable according to $\Delta$BIC. It now makes sense that we adopt the edge-on case as the orbital solution for this system since it shares similar statistical significance as the other models, but without taking inclination as one of the free parameters. We therefore employ the maximum likelihood result from the dynamical model as the best-fit orbital parameters for the rest of this work.

%%%%%%%%%%%%%%%%%%%%%%%%%%%%%%%%%%%%%%%%%%%%%%%%%%%%%%%%%%%%%%%%%%%%

\subsection{Stellar Activity}
\label{activity}

Periodic stellar activity masquerading as a planetary-like RV signal is a common false positive for RV planet detection. Many exoplanet discovery claims made using the RV method were later refuted due to presence of either short-term or long-term stellar activity cycles \citep{robertson2014a,robertson2014b,robertson2015b,kane2016a,lubin2021,simpson2022}. Thus it is often necessary to analyze the stellar activity indicators alongside the RVs to check for the validity of planetary RV signatures. If time series of stellar activity indicators produce similar periodicities as those in the RV data, or if the indicators are strongly correlated with the RVs, the planetary nature of the RV signal should be questioned. The host star HD~45364 was reported to be a non-active star with log$R^{\prime}_{HK}$~=~-4.94 and with a rotation period of $P_{rot}$~=~32 days \citep{correia2009}. However, no work has been done previously to check for the periodicities of activity indicators and correlation between activity signals and RVs. Here, since the majority of the RV data is from HARPS, we used stellar activity indicators associated with HARPS RVs for HD~45364. For each of the indicators: chromatic index (CRX), differential line width (dLW) \citep{zechmeister2018}, full-width-at-half-maximum (FWHM), and line contrast of the cross correlation function (CCF), we ran a periodogram search through Generalized Lomb Scargle periodogram (\texttt{GLS}; \citealp{zechmeister2009}) and \texttt{RVSearch} \citep{rosenthal2021} to double check for presence of periodic signals. Then, a correlation test was conducted using the Pearson correlation coefficient between each of the indicators and the individual RV contribution from each planet. No significant peaks above the 0.1\% FAP level were recovered in any of the activity indicators and no correlation were observed between activity indicators and each planetary RV contribution. These null results from the activity indicators confirm the two RV signals we see here are indeed due to planetary companions. 

We do note however that despite the absence of significant periodic stellar activity cycles, there is a noticeable trend present in CCF FWHM, CCF line contrast, and dLW. Line width shows a steady increase over time as indicated by FWHM and dLW whereas line contrast shows a decrease of line height over time. These indicators describe the spectral line shape change over time and the trend could be an indication of a longer period magnetic cycle. Such possibility could sometimes be verified by checking the CCF bisector span, or bisector inverse slope (BIS) that describes the overall change in line skewness caused by stellar magnetic activity \citep{queloz2001}. Unfortunately the HARPS BIS indicator contained for this target was poorly derived and was not usable for identifying activity as the source of the line shape change. Time series S-index or log$R^{\prime}_{HK}$ \citep{duncan1991} could also be used for such activity check yet they are unavailable from the HARPS public RV dataset. One other possible cause of the observed activity trend could be an additional low mass and dim stellar companion in the system, or a background star that is passing near HD~45364, potentially causing blending of some of the spectral lines and changing the line shape over time. However, no linear trend was observed in the RV data, either due to the stellar companion being too far away to cause noticeable RV trend, or the assumption of a stellar companion in the system is false. To further complicate the diagnosis, there were known instrument defocusing issues on both HARPS and HARPS-N that would affect the FWHM and contrast of the CCF to exhibit trend-like features \citep{locurto2015,benatti2017,barbato2019}. The exact cause of the changing in line shape is still unknown and the investigation is unfortunately beyond the scope of this work. Future RV observations that establish a longer baseline with the assistance of aforementioned activity indicators could hopefully resolve this mystery.

%%%%%%%%%%%%%%%%%%%%%%%%%%%%%%%%%%%%%%%%%%%%%%%%%%%%%%%%%%%%%%%%%%%%

\subsection{RV Search Completeness}
\label{completeness}

\begin{figure}[htbp!]
  \includegraphics[trim=40 30 30 30,clip,width=\columnwidth]{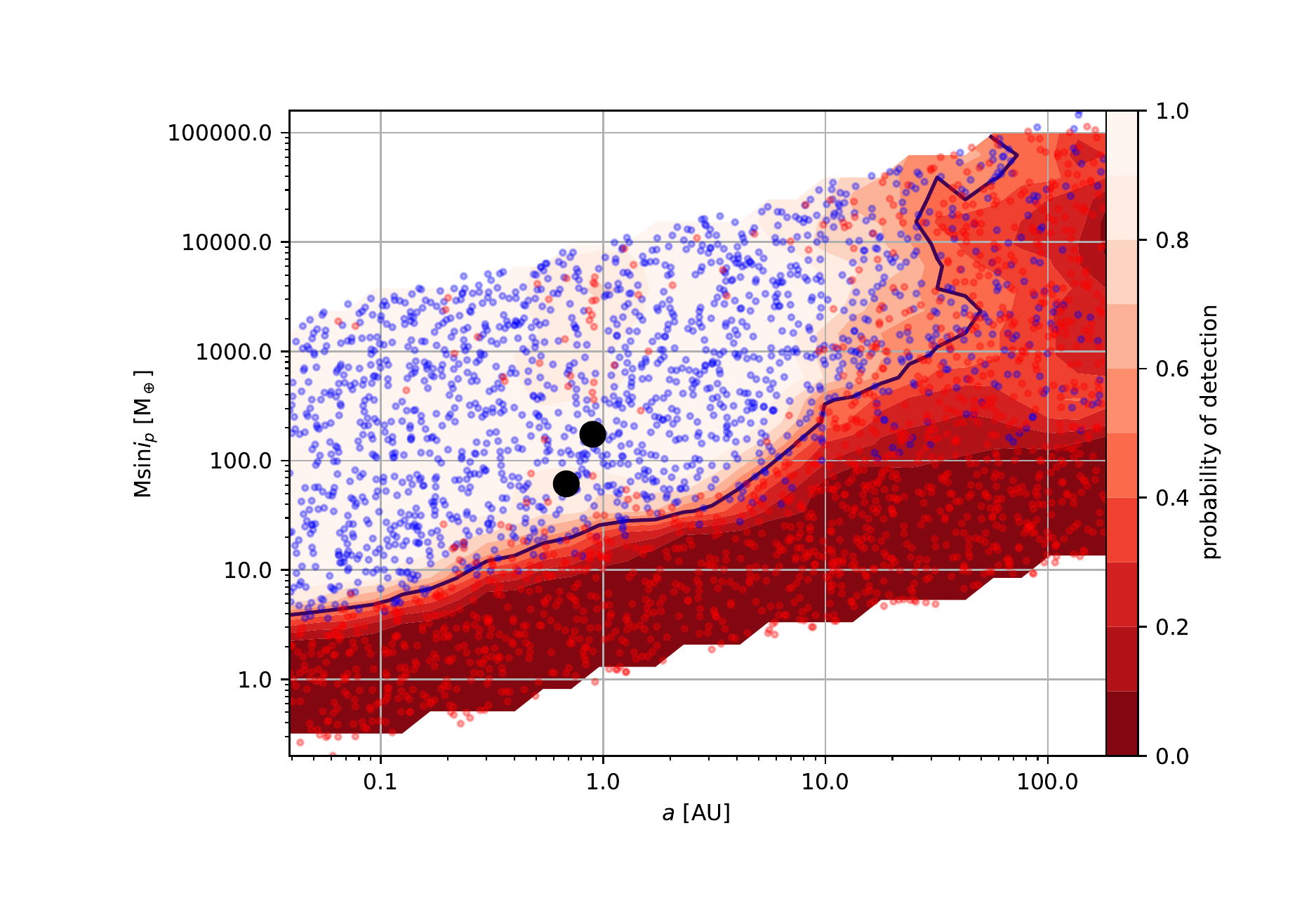}
  \caption{RV completeness contour of the HD~45364 dataset from the injection-recovery test. Two black dots are the known two planets in the system. Blue and red dots are the injected fictitious planet that was recovered and not recovered, respectively. The black line represents the 50\% detection probability of the contour. Detection probability was color coded according to the color bar on the right.}
  \label{fig:rvcompleteness}
\end{figure}

In addition to planetary discoveries, knowing the detection limit or sensitivity of the data is equally important because they entail what other bodies that could be detected but was not by the data, or what other objects could potentially be lurking in the system below the detection threshold \citep{wittenmyer2020a,li2021,rosenthal2021}. Such studies not only extend the knowledge of exoplanet population and system architecture, but also serve as a guideline to target selection of future space missions regarding the search for habitable terrestrial planets in the HZ for example, where the likelihood of the presence of such planets could determine the observing priorities of candidate systems. Here, we carried out an injection-recovery test within the $M_{p}$sin$i$ vs $a$ parameter space using \texttt{RVSearch} to determine the search completeness of our RV data. We injected 3000 fictitious planets into the current dataset and ran the iterative planet search within \texttt{RVSearch} in the same way as described in Section \ref{kepsol} to see whether the injected planet could be recovered. The fictitious planet was given an orbit with parameters drawn from log-uniform distributions for the planet's orbital period and minimum mass. The eccentricity was drawn from a Beta distribution following the result from \citet{kipping2013b}. The result of the injection-recovery test was then used to compute the RV search completeness contour which is shown in Figure \ref{fig:rvcompleteness}. The two known giant planets that have been the centerpiece of this paper are clearly above the detection threshold of our RV dataset. It is interesting to point out from this figure that, the data cannot rule out the possibility of additional very long orbital period companions since the current data is insensitive to objects, if there are any, at the very large separations from the host star. Thus a dim low mass stellar object that is gravitationally bound to the system is still a possible explanation to the observed trend in two of the activity indicators as we mentioned in the last subsection. In addition, the data suggests that additional gas giants with 1 Jupiter mass could be detected with 90\% detection probability under the current data sensitivity for separations up to $\sim$~7~au. However, the non-detection of such planets within our data set indicate the two we have now may be the only giant planets in the system within this separation range.

%%%%%%%%%%%%%%%%%%%%%%%%%%%%%%%%%%%%%%%%%%%%%%%%%%%%%%%%%%%%%%%%%%%%

\section{Dynamical State}
\label{dynamics}

\begin{figure}[htbp!]
  \includegraphics[trim=0 0 30 30,clip,width=\columnwidth]{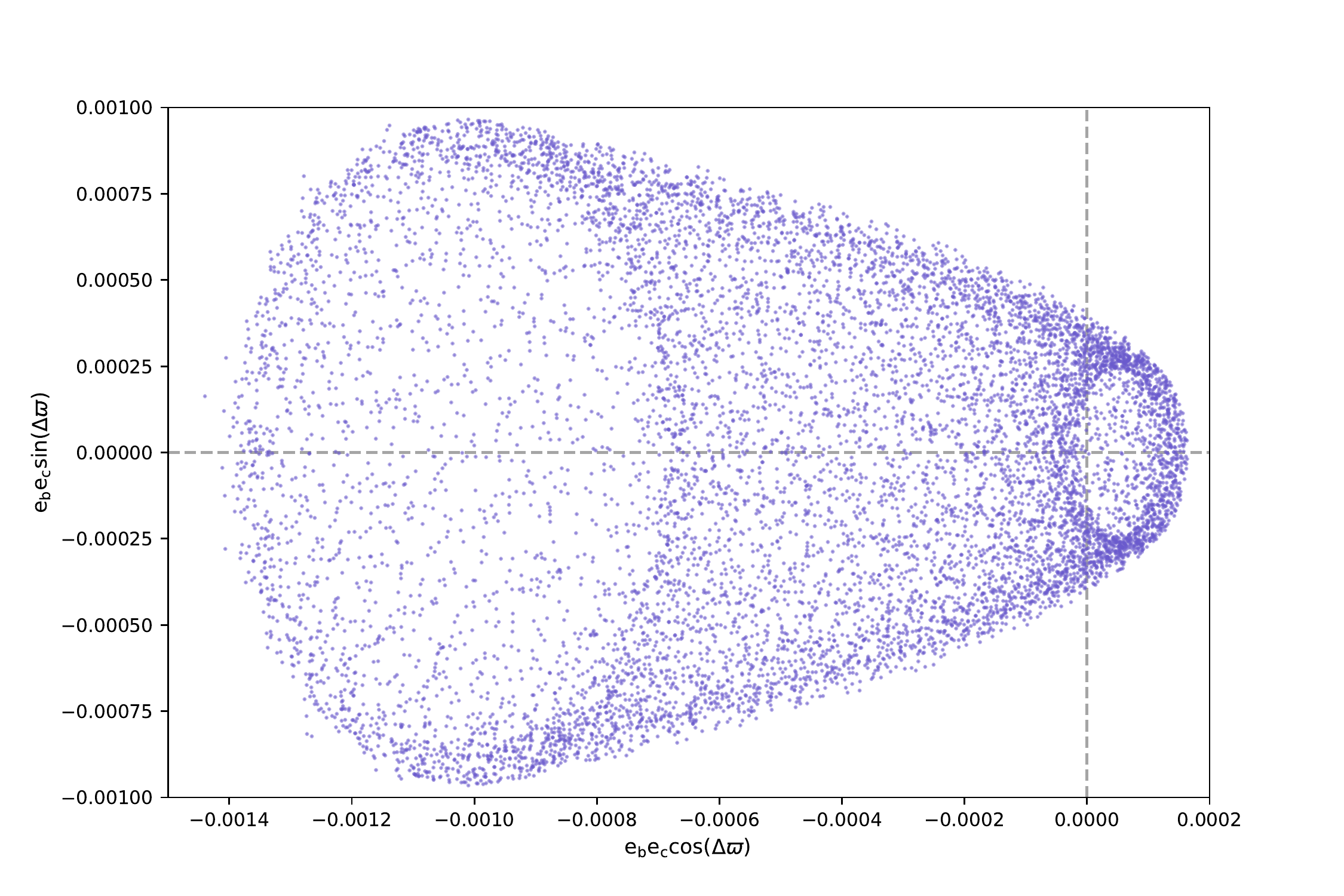}
  \caption{Polar plot of apsidal trajectory of the HD~45364 system for 10,000 years with each point in the plot being 1 year apart. Two apsidal modes (libration and circulation) are both present and the system's dynamics appears to be swinging in between the two.}
  \label{fig:polar}
\end{figure}

The new RV models, either Keplerian or dynamical, both yield slightly different orbital parameter values than those previously reported, especially for orbital periods, eccentricity, and planetary masses of the two giant planets. Combined with the fact that we do not see the predicted strong planet-planet interaction, the new picture leads us to wonder if the system could be in a different dynamical state. One way to study the dynamical behavior of the system is by calculating the apsidal trajectory and plotting the behavior in a polar plot of $e_{b}e_{c}$sin$\Delta\varpi$ against $e_{b}e_{c}$cos$\Delta\varpi$, where $\Delta\varpi$ = $\varpi_{b}$ - $\varpi_{c}$ (Figure~\ref{fig:polar}). Typically, if the trajectory encompasses the origin, the system is deemed to be circulating. If not, the system is librating and the type of libration depends on the location relative to the origin and the shape of the trajectory. These two modes of apsidal \textbf{behavior are} separated by a boundary called ``separatrix", for which the system could have a behavior of both libration and circulation if the trajectory comes close to the origin of the polar plot \citep{barnes2006a,barnes2006c}. In Figure~\ref{fig:polar}, the polar plot includes the result for 10,000 years with each plotted point being 1 year apart. Interestingly, the apsidal trajectory is showing two prominent oval shapes, one to the left of the origin, and the other one slightly to the right but encompasses the origin. These ovals may represent the two modes of apsidal behavior, antialigned libration (left oval) and circulation (right oval). However, the plot shows neither libration nor circulation seems to be in a stable state. Based on the number of data points that lie in between the two modes, the system does not appear to be confined in either one of the modes. This presents a possible but seemly uncommon scenario for the system's dynamical state in which neither libration nor circulation is dominating the overall behavior and the system is constantly seeking the balance between the two, thus forming a cone shape in the polar plot where the system is moving back to forth between the two modes. The feature seen in Figure~\ref{fig:polar} is still present if we extend the simulation duration to $10^7$~years.

\begin{figure}[htbp!]
  \includegraphics[trim=0 0 30 30,clip,width=\columnwidth]{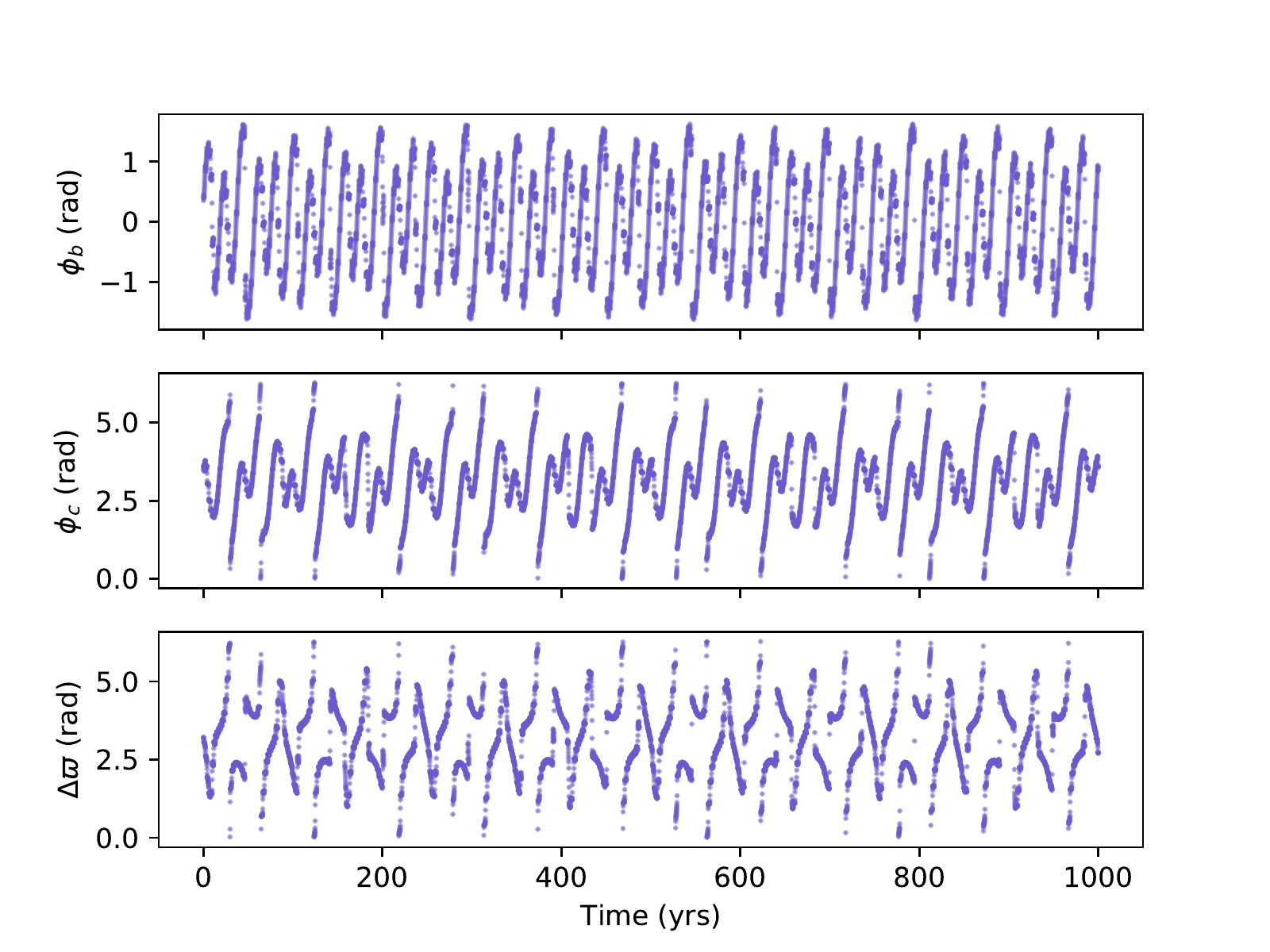}
  \caption{Evolution of the resonance angles $\phi_{b}$ (top panel), $\phi_{c}$ (middle panel), and $\Delta\varpi$ (bottom panel) for 1,000 years with a step of 0.1 year. The system exhibits both libration and circulation, with libration happen around 0 degree for $\phi_{b}$ and 180 degrees for $\phi_{c}$. The system goes through circulation, as indicated by $\phi_{c}$ and $\Delta\varpi$, near orbit alignment.}
  \label{fig:angle}
\end{figure}

\begin{figure}[htbp!]
  \includegraphics[trim=0 0 30 30,clip,width=\columnwidth]{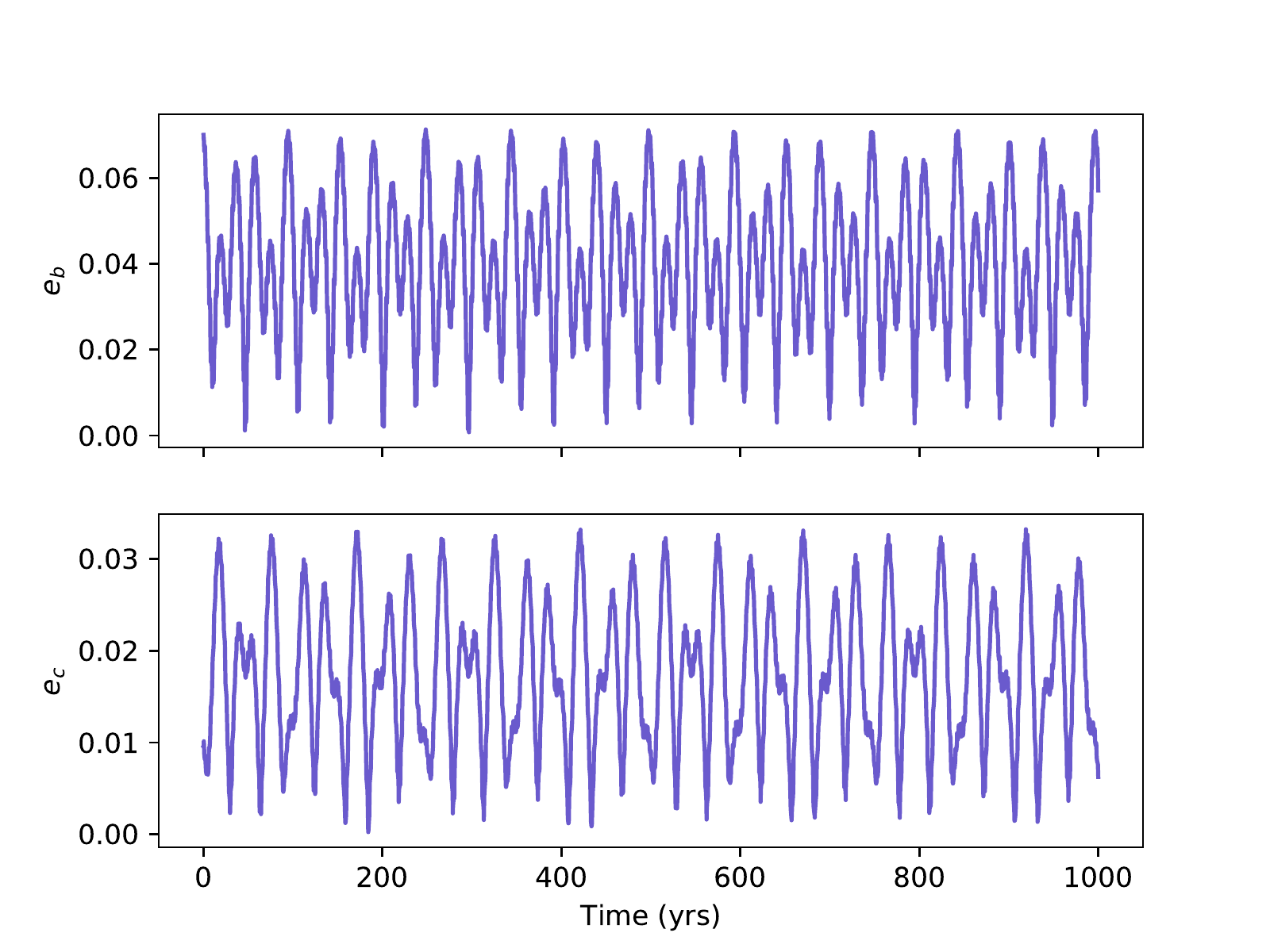}
  \caption{Eccentricity variations for planet b (upper panel) and c (lower panel) for 1,000 years. The variation goes through a similar $\sim$~250-year quasi-periodic cycle as those observed for the resonance angles in Figure~\ref{fig:angle}.}
  \label{fig:eccvar}
\end{figure}

The interesting dynamics seen in Figure~\ref{fig:polar} can be further investigated by calculating the evolution of resonance angles. For an interacting planet pair near a 3:2 MMR, the resonance angles can be calculated as:

\begin{equation}
\label{eqn:phi1}
    \phi_{b} = 3\lambda_{c} - 2\lambda_{b} - \varpi_{b}
\end{equation}

\begin{equation}
\label{eqn:phi2}
    \phi_{c} = 3\lambda_{c} - 2\lambda_{b} - \varpi_{c}
\end{equation}
We ran a short integration of 1,000 years and recorded the resonance angles every 0.1 years. The evolution of both angles and $\Delta\varpi$ is shown in Figure~\ref{fig:angle}. Indeed, the top panel of Figure~\ref{fig:angle} shows that $\phi_{b}$ is librating around 0$^{\circ}$ whereas $\phi_{c}$ seems to be stuck in between circulation and libration, with the libration pattern around 180$^{\circ}$. Upon closer examination, $\phi_{c}$ exhibits several similar but not the same libration and circulation cycles every $\sim$~250 years. Within each quasi-period, $\phi_{c}$ goes through mixed periods of libration in between each circulation. Such ``nodding" behavior where the resonance angle goes through phases of libration and circulation repeatedly over time was studied in detail in \citet{ketchum2013} and appears to be related to the perturber's orbital eccentricity variation. This is reflected in the eccentricity variation of both planet b and c in Figure~\ref{fig:eccvar} where the duration and time step are the same as those for Figure~\ref{fig:angle}. The b and c planet go through eccentricity variation of $\sim$~0 to 0.07 and $\sim$~0 to 0.03, respectively, and we verified these variations last for simulations of $10^7$~years. Looking closely at both eccentricities in Figure~\ref{fig:eccvar}, both planets are going through a similar quasi-periodic cycles of eccentricity variation every $\sim$~250 years, suggesting the orbital shape and orientation change could be the cause of circulation we see in Figure~\ref{fig:polar} and \ref{fig:angle}. Since circulations happen when the planet pair's orbits are close to alignment ($\Delta\varpi$~=~0) according to Figure~\ref{fig:angle}, this suggests $\phi_{c}$ goes through circulation when the interaction between the two planets is at a minimum. Due to the eccentricity variation, $\phi_{c}$ cycles back into libration when orbits are not aligned and planets undergo interaction with each other for a period of time, until the next orbit alignment and circulation. 

The libration patterns we see here in Figure~\ref{fig:angle} when the system goes through libration phase are largely similar to those in previous works (see for example Figure~9 in R10), except that both resonance angles in this case are librating with a much larger amplitude, especially in the case of $\phi_{c}$ where libration becomes so large that sometimes it goes into circulation regime. Both larger libration amplitude and two apsidal modes one after another indicate that instead of being within a 3:2 MMR, the planet pair is moving in and out of the resonance repeatedly, in a ``quasi-resonance" state. Such type of motion between libration and circulation near the ``separatrix" concerning the HD~45364 system was first suggested in C13 regarding the predicted orbital solution under the type III migration mechanism proposed by R10 and also one of the solutions under the type I, type II combination mechanism by C13, although not directly observed in their models. Given the similarity of the system's dynamical behavior between our model and some of those from R10 and C13, we therefore conclude that the system is indeed in a ``quasi-resonance" dynamical state and any migration mechanisms that predicted near circular orbit solutions should be preferred as the formation pathway to the configuration that we see today in HD~45364.

%%%%%%%%%%%%%%%%%%%%%%%%%%%%%%%%%%%%%%%%%%%%%%%%%%%%%%%%%%%%%%%%%%%%

\section{Habitability}
\label{habitability}

The previous orbital configuration with two planets strongly interacting on mildly eccentric orbits may present a challenge to allowing potentially habitable locations within the system. However, the revised orbital solution for the system with near circular orbits (right panel of Figure~\ref{fig:hzorbits}) and, more importantly, the minimal gravitational interaction between the two giant planets, as shown in previous sections, open up such a possibility. Here, we briefly discuss potential habitability aspects of the system, including the prospect of terrestrial planets within the HZ and moons harbored by the known giant planets.

%%%%%%%%%%%%%%%%%%%%%%%%%%%%%%%%%%%%%%%%%%%%%%%%%%%%%%%%%%%%%%%%%%%%

\subsection{Additional Planets}
\label{planets}

Giant planets observed to orbit within the HZ of the system are believed to have formed further out beyond the snow line, then gradually migrated inwards through planet-disk interaction. Such migration history may present a significant challenge to the formation of terrestrial planets within the HZ because of the constant gravitational perturbation to the nearby material in the disk, inhibiting the material buildup that would have eventually formed other planets. However, past works have shown that Earth-mass terrestrial planets could indeed survive the giant planet migration and start the formation process within the HZ during or after giant planet's passage of the HZ \citep{raymond2006a,raymond2006d}. Here, we conducted a dynamical simulation to test for the viable locations where another Earth-mass terrestrial planet could be present.

\begin{figure*}[htbp!]
  \includegraphics[trim=50 0 50 30,clip,width=\textwidth]{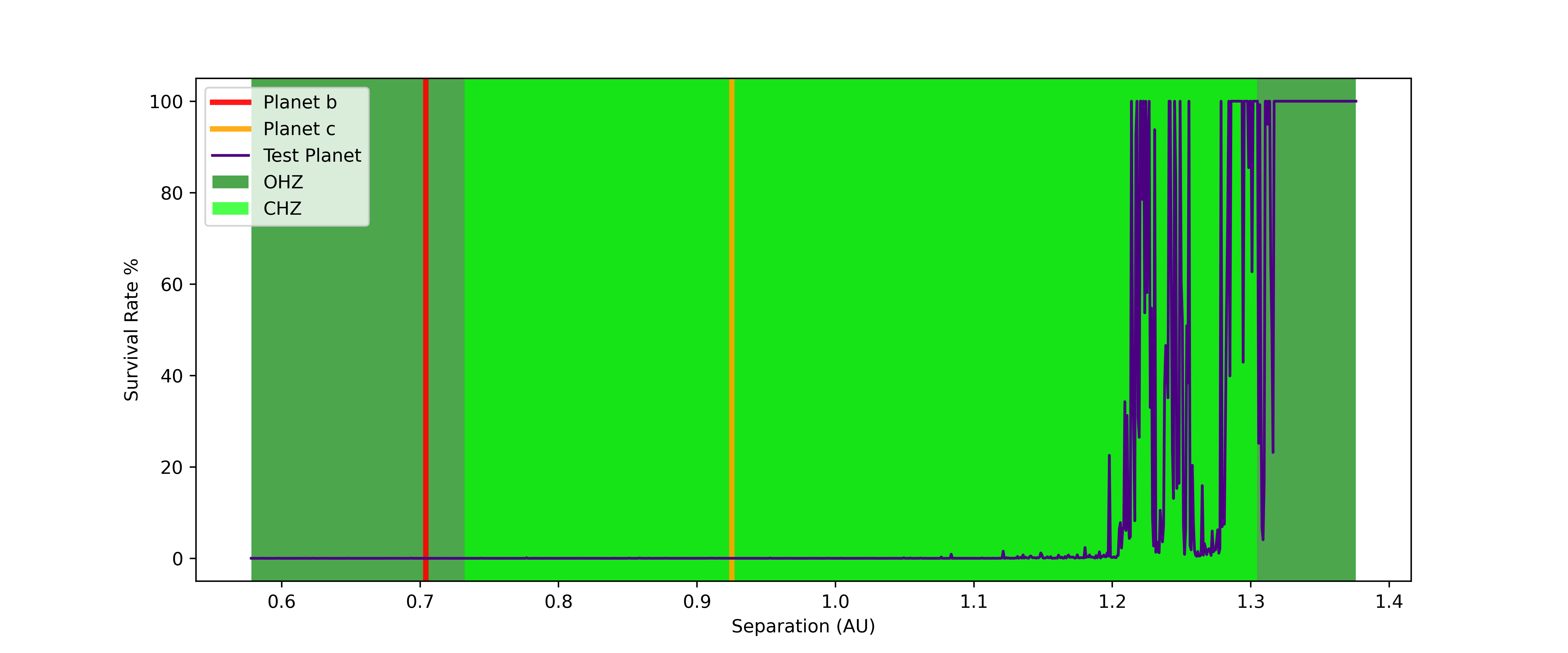}
  \caption{Simulation result for a test particle with an Earth mass planet injected at 1000 different locations evenly spaced between the boundaries of OHZ. Particle survival rates were calculated as the percentage of the total integration time that the particle survived and remained stable within the system. Only at separations where the particle achieved 100\% survival rate do we consider the location dynamically viable for the presence of an additional Earth mass terrestrial planet. Such locations are only possible near the outer edge of the HZ.}
  \label{fig:survivalrate}
\end{figure*}

We made use of the \texttt{REBOUND} package to carry out the dynamical simulation. For the two known giant planets in the system, we assumed an edge-on orbital configuration derived from our dynamical model presented earlier (Table~\ref{tab:param}). We then injected an Earth-mass planet in a circular orbit into the existing system at 1000 different locations, one location at a time, evenly spaced within the range of the OHZ from 0.578~au to 1.376~au (see Section~\ref{system}). We used the WHFast integrator with a fixed time step of $\sim$~8 days, roughly 1/20th the orbital period of a test particle orbiting at the inner edge of the OHZ. The system with the hypothetical Earth-mass planet was integrated for $10^7$~years to test for system stability and the survival of the terrestrial planet, similar to the methodology described by \citet{kane2015a,kane2015b,kane2022a}. Results were recorded every 100 years and the simulation for each test location would stop if any one of the three planets were ejected from the system. After the simulation, we computed the survival rate of the test planet at each test location, defined as the percentage amount of time the test planet would remain in the system throughout the entire integration duration. Any locations with the Earth-mass planet having 100\% survival rate are considered dynamically viable locations. The simulation results are shown in Figure~\ref{fig:survivalrate}. As expected, most of the locations within the HZ are prohibitive to the presence of an Earth-mass terrestrial planet due to the influence of the two giant planets within the HZ, as indicated by the red and orange vertical lines in Figure~\ref{fig:survivalrate}. In the vicinity of the two known planets, the injected planet had near zero survival rates from the inner edge of the HZ up to $\sim$~1.2~au, meaning that the small planet was ejected momentarily after the start of the integration. However, as the test planet was moved further away from the influence of the two giants and towards the outer edge of the HZ, the dynamically stable area opened up for the terrestrial planet and it was able to achieve full survival at the end of the simulation at locations near the outer edge of the HZ, some even within the outer boundaries of the CHZ. These results show that, provided the migration of the giant planets allow subsequent formation of terrestrial planets within the system, those planets may retain orbits within the outer edge of the HZ.

%%%%%%%%%%%%%%%%%%%%%%%%%%%%%%%%%%%%%%%%%%%%%%%%%%%%%%%%%%%%%%%%%%%%

\subsection{Implications for Exomoons}
\label{exomoons}

Giant planets orbiting within the HZ of the system are now increasingly scrutinized for their prospects as hosts of potential habitability. These planets themselves may not have the environment suitable for habitability studies, but potential exomoons orbiting around these giant planets might do thanks to the combined global energy budget from the star, the planet and tidal energy \citep{tinney2011,heller2014c,hill2018}. Significant studies have been carried out regarding orbital stability of moons \citep{barnes2002,kipping2009b,gong2013,kane2017c,hong2018,quarles2020,dobos2021}, especially those around planets that have experienced migration \citep{spalding2016}. For HD~45364, a detailed investigation of the formation and dynamics of planet-moon systems is beyond the scope of this work. We can, however, provide an estimate of the Hill radius of both giant planets and the critical orbits of potential exomoons, provided by the following equations:

\begin{equation}
\label{eqn:hill}
    r_{H} \approx a(1-e)(\frac{M_{p}}{3M_{\star}})^{1/3}
\end{equation}

\begin{equation}
\label{eqn:crit}
    a_\mathrm{crit} \approx fr_{H}
\end{equation}

According to Equation~\ref{eqn:hill} above, and using parameters from the best dynamical fit from Table~\ref{tab:param}, the size of planet b and c's Hill radius are $r_{H,b}$~$\approx$~0.026~au and $r_{H,c}$~$\approx$~0.053~au, respectively. However, the calculation of Hill radius is only an estimation that does not take into account any other external perturbations \citep{kipping2009b}. Following \citet{barnes2002} and \citet{kipping2009b}, we adopt $f$~=~1/3 and use $a_\mathrm{crit}$ from Equation~\ref{eqn:crit} as a conservative outer boundary for the possible locations of exomoons around exoplanets. For the b planet, we estimated $a_\mathrm{crit,b}$~$\approx$~0.009~au, and for the c planet, $a_{crit,c}$~$\approx$~0.018~au. Despite the best estimate, this upper bound estimate should not be mistaken as the limit below which an exomoon, if there exists one, is guaranteed to be stable around the planet it orbits. As pointed out by \citet{spalding2016}, survival of exomoons from type II migration of giant planets depends on the location the planets was migrating from. However, if there are multiple moons orbiting around a giant planet in an MMR configuration, such moon destruction through ``evection resonance" may be quenched due to moon-moon interactions \citep{spalding2016}.

%%%%%%%%%%%%%%%%%%%%%%%%%%%%%%%%%%%%%%%%%%%%%%%%%%%%%%%%%%%%%%%%%%%%

\section{Discussion and Conclusions}
\label{conclusions}

The orbital solution published by C09 in 2009 resulted in two giant planets orbiting in the HZ of the system with mildly eccentric orbits. The planet pair was found to reside in a 3:2 MMR that prevent the planets from experiencing close encounters with each other. Several works have since then attempted to reconstruct the formation pathway to the observed configuration at the time through different planetary migration models. R10 proposed a type III migration model and predicted near circular orbits for both planets rather than eccentric ones. C13 instead proposed a combination of type I and type II migration scenario and predicted both eccentric and circular orbits are possible under different assumptions. All these works have stressed the importance of future RVs in helping resolve the model degeneracy. 

In this work, we revisited the system 13 years later using a substantially improved RV dataset. We presented a new orbital solution for the two planets within the HD~45364 system using new RV data from HARPS and HIRES. The latest RVs allowed us to extend the observational baseline to almost 18 years, almost four times the baseline used by the original publication by C09 13 years ago. We conducted a reanalysis of the RV data through both Keplerian and dynamical models. Both models point to near circular orbits for the two planets instead of having mild eccentricities. In addition, small changes were observed for the orbital periods and planetary masses. As a consequence, the orbits are now more separated, and interactions between the planets may not be as strong as expected. Indeed, by comparing our Keplerian and dynamical models, we did not observe the predicted large amplitude of $\Delta$RV over time between the two models. The $\Delta$RV we determined was consistent with the noise level from both the star and the instruments. We thus conclude that there is negligible planet-planet interactions within the sensitivity of our RV dataset. We also attempted to determine the orbital inclination of the system through dynamical fitting. However, due to much weaker interactions between the planet pair, orbital inclination can only be constrained to $\ge$~40${^\circ}$ and edge-on case is so far preferred.

The orbital dynamics was studied in detail given the change in the orbital configuration of the system and planetary masses. We integrated the system for $10^7$~years and tracked the evolution of orbital elements for both planets. By tracking the apsidal trajectory and resonance angle evolution, we found that the system was not quite in a 3:2 MMR. Rather, the system exhibits both libration and circulation, and thus appears to be moving in and out of the 3:2 MMR. The libration and circulation pattern is in a quasi-periodic cycle that is consistent with a similar cycle observed in both planet's eccentricity variations, suggesting the observed new dynamical behavior of the system is related to the periodic change in the magnitude of interaction between the two planets. Therefore, HD~45364, the first exoplanetary system discovered to host a 3:2 MMR, is actually in a quasi-resonance state, rather than being in a true resonance state. The result of our new RV models and dynamical analysis successfully confirmed predictions from previous works. More importantly, our result suggests that any migration models that predict near circular orbits for both planets should be the preferred migration scenarios for the HD~45364 system. This work demonstrates the importance of continued RV monitoring of a system and how it could impact our understanding of the system's dynamics.

Our orbital solution indicates the b planet resides within the inner OHZ, the c planet resides within the CHZ, and the orbits of the two planets no longer experience signs of a close encounter (Figure~\ref{fig:hzorbits}). The new system architecture opens up possible habitable locations in the HZ, and we investigated the habitability prospects of the system. An Earth-mass terrestrial planet was injected into the HZ and tested for dynamically viable locations near the two giant planets. The simulation indicated such a planet is indeed possible near the outer edge of the HZ, and detection of such a planet would be challenging since the expected RV amplitude would be well below the current RV sensitivity (e.g. $K$~$\approx$~8.5~cm~s$^{-1}$ at 1.35~au). Exomoons around the two giant planets present another possibility for potentially habitable locations within the system, based on the simple estimate of the Hill radius of the two planets. If future observations could detect such objects, the new piece of data could provide further insight into the rich dynamical history that HD~45364 presents. This system therefore holds great value for future space-based missions that searches for potentially habitable bodies in nearby exoplanetary systems.

There remains significant further work regarding the HD~45364 system. Transit observations could reveal more about this system. Based on the stellar radius value from Section~\ref{system} and orbital parameters of the two planets from Table~\ref{tab:param}, the transit probability is only 0.61\% for the b planet and 0.46\% for the c planet, assuming randomly oriented orbits. Assuming planetary radii of $R_{b}~=~0.81$~$R_{\rm J}$ and $R_{c}~=~1.22$~$R_{\rm J}$, derived from the best-fit mass values using the mass-radius relationship of \citet{chen2017}, the two planets would have transit depths of 0.87\% and 1.98\%, respectively. HD~45364 was observed by \textit{TESS} in sectors 6 and 7 from 2,458,468 to 2,458,516 BJD. According to our best dynamical model fit, inferior conjunction times are $T_{c,b}$~=~$2,458,453^{+23}_{-24}$ BJD and $T_{c,c}$~=~$2,458,487^{+53}_{-66}$ BJD for planet b and c, respectively. The estimated conjunction times of both planets sit very close to the \textit{TESS} observation window. We checked \textit{TESS} data and no transit was detected for impact parameter b~$<$~1. The transit non-detection suggests that either \textit{TESS} observations had a near miss of the event in time or the system is not edge-on. Given the large transit depth of both planets, ground-based follow-up observations could help confirm or rule out the transit scenarios. Future transit observations could potentially disclose more information about this system, such as planetary radius, atmosphere, transit timing variations, and even additional terrestrial planet and exomoon detections, if the system is edge-on. In addition, continued RV monitoring by the precision RV facilities could further help refine the orbital solutions, track dynamics of the system, and investigate the source of the mysterious long term trend seen in some of the activity indicators to determine whether it is of companion or activity origin. HD~45364 is without a doubt one of the most interesting systems in many areas of exoplanet study.

%%%%%%%%%%%%%%%%%%%%%%%%%%%%%%%%%%%%%%%%%%%%%%%%%%%%%%%%%%%%%%%%%%%%

\section*{Acknowledgements}

The authors wish to thank Hanno Rein for discussion and suggestions regarding the use of the \texttt{REBOUND} package. The authors also would like to thank Rory Barnes for conversation regarding the system's dynamics. We thank the anonymous referee for the valuable comments that greatly improve the presentation of this work. P.D. acknowledges support from a 51 Pegasi b Postdoctoral Fellowship from the Heising-Simons Foundation. Dynamical simulations in this paper made use of the \texttt{REBOUND} code which is freely available at \url{http://github.com/hannorein/rebound}. This research has made use of the NASA Exoplanet Archive, which is operated by the California Institute of Technology, under contract with the National Aeronautics and Space Administration under the Exoplanet Exploration Program. This work has also made use of The Habitable Zone Gallery at \url{hzgallery.org} \citep{kane2012a}. The results reported herein benefited from collaborations and/or information exchange within NASA's Nexus for Exoplanet System Science (NExSS) research coordination network sponsored by NASA's Science Mission Directorate.

%%%%%%%%%%%%%%%%%%%%%%%%%%%%%%%%%%%%%%%%%%%%%%%%%%%%%%%%%%%%%%%%%%%%

\software{\texttt{GLS} \citep{zechmeister2009}, \texttt{RadVel} \citep{fulton2018a}, \texttt{REBOUND} \citep{rein2012a}, \texttt{RVSearch} \citep{rosenthal2021}}, \texttt{RVStab} \citep{rosenthal2019}

%%%%%%%%%%%%%%%%%%%%%%%%%%%%%%%%%%%%%%%%%%%%%%%%%%%%%%%%%%%%%%%%%%%%

%\bibliographystyle{aasjournal}
%\bibliography{references}

%%%%%%%%%%%%%%%%%%%%%%%%%%%%%%%%%%%%%%%%%%%%%%%%%%%%%%%%%%%%%%%%%%%%

\end{document}